\documentclass[10pt]{iopart}
\usepackage{amssymb}
\usepackage{amsopn}
\usepackage{graphics}
\usepackage{graphicx}
\usepackage[dvips]{hyperref}
\usepackage{subfigure}
\usepackage{psfrag}
\usepackage[dvips]{color}
\newcommand{\dd}[0]{\mathrm{d}}
\newcommand{\be}[0]{\begin{eqnarray}}
\newcommand{\ee}[0]{\end{eqnarray}}
\newcommand{\nn}{\nonumber}
\newcommand{\fd}{\frac{1}{2}}

\newcommand{\tq}{\theta_{q+\frac{1}{2}}}
\newcommand{\Hs}{H_\mathrm{s}}
\newcommand{\raps}{\zeta_\mathrm{s}}
\newcommand{\tc}{t_\mathrm{c}}
\newcommand{\sint}{\sigma_{\mathrm{int}}}
\begin{document}
\title[]{Boundary field induced first order transition in
the 2D Ising model: Exact study}

\author{Maxime Clusel$^{\dag} $ and  Jean-Yves Fortin$^\ddag$}
\address{\dag\ Institut Laue-Langevin, 6 rue Horowitz BP156 X, 38042 Grenoble
cedex, France}
\address{\ddag\
Laboratoire Poncelet, 119002, Bolshoy Vlasyevskiy Pereulok 11,
Moscow, Russia}
\ead{clusel@ill.fr, fortin@lpt1.u-strasbg.fr}

\begin{abstract}
We present in this article an exact study of a first order
transition induced by an inhomogeneous boundary
magnetic field in the 2D Ising model. From a previous analysis of the
interfacial free energy in the discrete case
(\JPA {\bf 38}, 2849, 2005) we identify, using an asymptotic expansion in the
thermodynamic limit,
the line of transition that separates the regime where the interface is localised
near the boundary from the one
where it is propagating inside the bulk. In particular, the critical line has a
strong dependence on the aspect ratio of the lattice.
\end{abstract}
\pacs{02.30.Ik ; 05.50.+q ; 05.70.Fh}
\submitto{\JPA}

\maketitle
%
\section{Introduction}

The 2D Ising model is certainly one of the most famous statistical
models in physics with connections in various fields of research.
Despite its apparent simplicity, it contains generic properties
for phase transitions. From the interpretation of the
Ising model as a lattice gas model, it was used as a simple model
for wetting transitions. A number of theoretical and numerical tools
have been developed to study such transitions. One of the first
steps comes from McCoy and Wu who developed the theory of Toeplitz
determinants~\cite{mccoy67b} in order to solve the 2D Ising model
with uniform boundary magnetic field (see also \cite{mccoybook,wu66,mccoy67}.
Surface fields can
be considered as the effect of boundary impurities or constraints that change 
the properties of bulk spins. They also can be viewed as a chemical potential difference
between the bulk and the wall of the system
 in a binary mixture. The wall can indeed attract one species more than the
other, leading so to an effective field~\cite{gennes78}.
 A major contribution is due to D.B. Abraham in the early 80's
who developed efficient methods~\cite{abraham.74} based on transfer matrix
techniques in order to solve the 2D Ising model with various fixed boundary
conditions (the spins are constraint
to be up or down on the boundary according to different profiles of the
boundary spins). Theses techniques allow the computation of the
surface tension energy for different profiles of the boundary conditions.
 Important questions arise whether the interface between two phases
and produced by boundary impurities is diffuse or sharp (see discussion
in~\cite{abraham80})
and how it diverges with the system size. These questions are closely related in
general to the
existence of roughening transitions in such systems close to the bulk critical
temperature.
 A generalisation of McCoy and Wu result for the uniform boundary field case is
to
 take two opposite surface fields $H_1$ and $H_2$ on a Ising strip, where a
domain
 wall propagates in the middle of the strip when the fields have opposite
signs~\cite{abraham80,yangfisher80,maciolek96}. A
 wetting temperature $T_\mathrm{w}(H_1,H_2)$ occurs at slightly below the bulk
temperature $T_\mathrm{c}$  where a
 well defined interface appears in the middle of the strip. This temperature
can
 also be defined as the transition between diffuse and sharp interfacial 
regimes.
 In the particular case of an infinite strip,
 the width of the strip is only important for the magnetisation profile and for
the scaling of the typical
 length of the interface
extension~\cite{maciolek96,abraham82,abraham84,abraham88}, which diverges
typically like the inverse of $(T-T_\mathrm{w})$ above $T_\mathrm{w}$, giving an interface correlation
length exponent equal to unity. Technically, a generic way to implement a 
finite boundary magnetic field from an infinite one is to take the bonds 
perpendicular
  to the surface field line a fraction of the bulk bonds. For example if the
spins subject to an infinite magnetic field $H_1=\pm \infty$ and
  located along the surface are noted $\sigma_{1,n}$, $n=1\dots L_y$, and
if the bulk coupling is $J$, taking $J_0$ as the coupling between $\sigma_{1,n}$
and $\sigma_{2,n}$
 generates on spins $\sigma_{2,n}$ a field proportional to $\pm J_0/J$. The
value of this field
 can then be tuned by varying $J_0$, and this specificity is discussed in
several publications~\cite{abraham80,abraham88,abraham81}.
 In these different cases, the way the thermodynamic limit is taken is also
important.\\
 The effect of an inhomogeneous magnetic field on one border is therefore an 
interesting
problem because it generates a localised interface whose extension
 inside the bulk can be studied in the framework of diffusion processes. In 
particular the interface generated by two opposite magnetic fields of infinite 
amplitude in the critical Ising model is a typical example of Schramm-Loewner 
evolutions (SLE) \cite{cardy05}. It is generally restricted to the conditions of infinite strip models, and the typical size of the interface is scaled with the strip width.
  In this paper we want to focus on another particular geometry that can be
treated exactly, where we apply a non homogeneous surface field on a rectangular Ising  lattice. We will see that the interfacial free energy can be evaluated in 
the discrete case and that, in the 
 thermodynamic limit, two dominant and competing terms determine whether the 
interface stays localised on the surface or is extended inside the bulk, 
depending on the lattice aspect ratio.
This model, even if the perturbation stays on the surface, 
shows some interesting complexity in the bulk. The 
 techniques used here are based on Grassmann 
variables~\cite{plechko85,plechko88,samuel,bugrij90,nojima,jpa05} and can
 interestingly be extended to other configurations as well. Experimentally it 
might be important to take account of the geometry of the system like the ratio 
of typical length scales to determine the values of the critical field or 
temperature, as we are going to see in the following.\\
 This article is organised as follow: We summarise in section~\ref{model} a 
previous exact result in the discrete case of the free energy contribution due 
to a inhomogeneous magnetic field. The analytical expression
 is however not easy to deal with and is rather obscure from a physical point of view, even if it has the advantage to prevent the use of 
 any cut-off, inherent to continuous models. Simple arguments for the formation 
of the bulk interface as function of field amplitude are given at zero 
temperature. An asymptotic method to obtain the thermodynamic  limit of the 
discrete solution is presented in section~\ref{analysis}, and show 
the simple effect of the system size ratio on the transition line. Monte Carlo 
simulations are also used to confirm the phase diagram. We then conclude and 
propose briefly the possibility of extensions of our method to other surface effects.
  
\section{The model}\label{model}
\begin{figure}[]
\begin{center}
\includegraphics[angle=-90,width=0.8\linewidth]{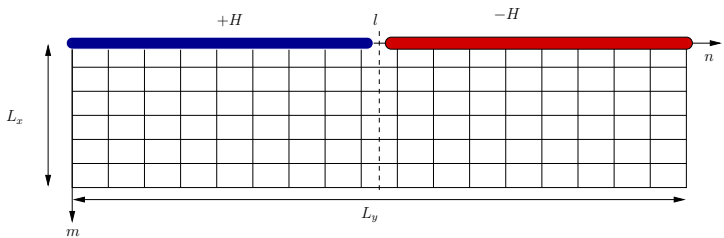}
\caption{\label{lattice}Inhomogeneous field configuration on the lattice. The 
field is positive for the sites $(m,n)=(1,1)\dots (1,L_y/2)$ and
negative on the rest of the line. Periodic conditions are imposed along the 
$y$-direction, and open conditions on the transverse
direction.}
\end{center}
\end{figure}
\begin{figure}
\begin{center}
\includegraphics[scale=0.6]{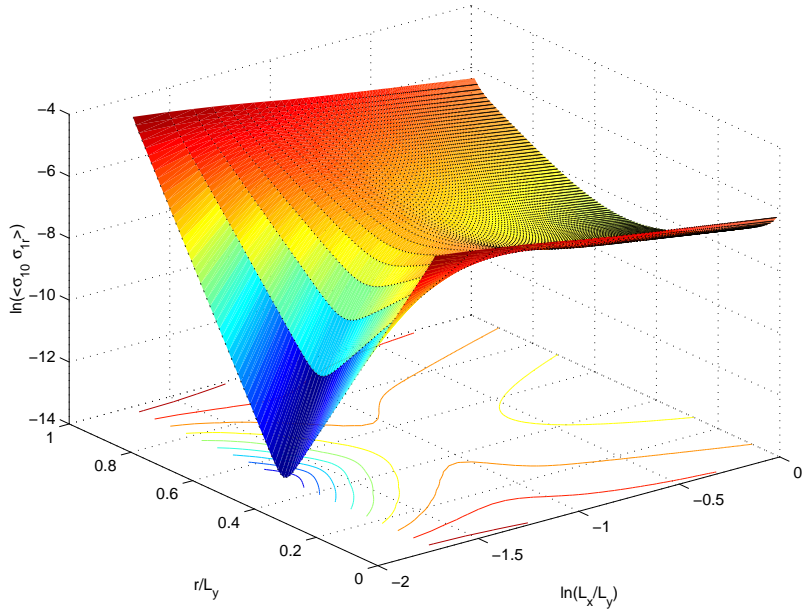}
\caption{\label{crossover}Boundary spin-spin correlation function $\langle 
\sigma_{10}\sigma_{1r}\rangle$ as a function of $r/L_y$ and aspect ratio 
$\zeta=L_x/L_y$, at $T=2J$, $H=0.1J$ and for $L_y=100$. Note the crossover from 
a 1D behaviour to a 2D behaviour at $\zeta \simeq 1/4$.}
\end{center}
\end{figure}

 We consider in this paper a finite 2D Ising system with a non homogeneous
magnetic field $h_n$ located on one boundary of the system. This system
is periodic along the $y$-direction, see figure \ref{lattice}, and with open
boundaries for the transverse $x$-direction, one of the latter boundaries being 
under a magnetic field. The Hamiltonian is simply given by 
\be \label{Ham}
\mathcal{H}=-J\sum_{m,n=1}^{L_x,L_y}(\sigma_{m n}\sigma_{m+1 n}+ \sigma_{m n}
\sigma_{m n+1})-\sum_{n=1}^{L_y}h_n\sigma_{1n},
\ee
with $\sigma_{m 1}=\sigma_{m L_y+1}$ and $\sigma_{0 n}=\sigma_{L_x+1 n}=0$.
The notations are the same as in a previous publication~\cite{jpa05} where 
an exact expression for the free energy in the case of finite size and discrete lattice 
case was obtained using Grassmann techniques and Plechko method based on 
Grassmann operator ordering~\cite{plechko85,plechko88}. These
operators replace basically spin operators, and in the 2D case, they lead to 
a Grassmannian quadratic action which is exactly solvable in the Fourier space.  
We have found in particular an exact ordering of border operators associated 
with a general boundary magnetic field and then obtained, after integration over 
bulk degrees of freedom, a quadratic 1D action with effective Grassmannian 
magnetic fields  (see equation (61) in reference~\cite{jpa05}),
for equal system sizes $L_x=L_y$. This 1D action represents the free energy 
contribution from the boundary fields. 
The case with different sizes $L_x\neq L_y$ is easily implemented as we will see 
below. As an application, we have considered in~\cite{jpa05} the interface initiated
by an inhomogeneous magnetic field of configuration shown in figure 
\ref{lattice}, with $h_n=H$ for $n=1\dots L_y/2$ and $h_n=-H$ for 
$n=L_y/2+1\dots L_y$. Computing the interface energy $\sint$ in the 
inhomogeneous case is equivalent to solve a set of Grassmannian two points 
correlation functions (see equation (74) in~\cite{jpa05}) using the previous 1D 
boundary action for the homogeneous case, which is done exactly. For more general configurations, with 
different sequences of fields $(H_k;l_k)$, \textit{i.e.} $h_n=H_k$ for 
$n=l_{k}+1\dots l_{k+1}$, the problem is treated identically~\cite{jpa05}.
  In this paper, we will consider the solution for the interface energy 
(see equation (84) in~\cite{jpa05} and below), with $L_x\ne L_y$,
  and study the thermodynamic limit with fixed aspect ratio $\zeta=L_x/L_y$. 
Numerically, the difficulty to compute this free energy arises from the fact 
that it is expressed as the logarithm of some argument which is an exponential 
small number in the system size, especially at low temperature. 
Moreover, to study in detail the phase transition corresponding to the 
propagation of the
 interface from where the magnetic field changes its sign, we need to take 
directly the 
 thermodynamic limit, and the purpose of this paper is precisely to study how 
the discrete expression of the interface energy behaves when we take the limits 
 $L_x,L_y\rightarrow \infty$.  \\
 \begin{figure}[]
\begin{center}
\includegraphics[angle=0,scale=0.4]{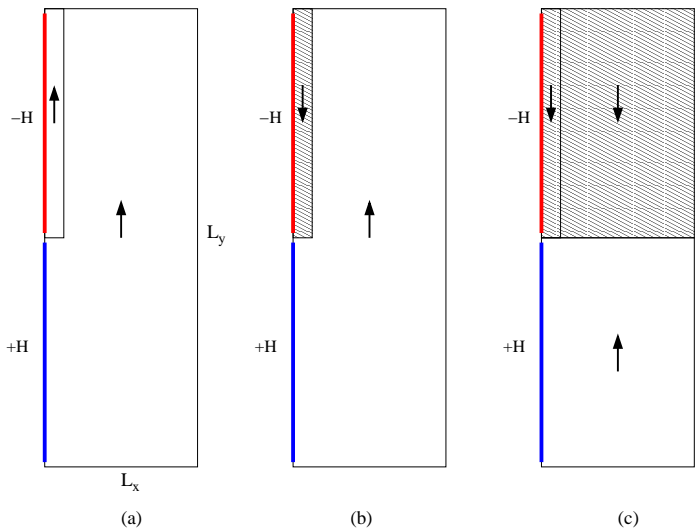}
\caption{\label{configurations}Spins configurations at $T=0$ under various 
conditions: (a) $L_x/L_y>\raps$, $H<J(1+4/L_y)$, (b) $L_x/L_y>\raps$, $H>J(1+4/L_y)$ and (c), 
$L_x/L_y<\raps$, $H>\Hs=4JL_x/L_y$.}
\end{center}
\end{figure}
  The presence of an interface phase transition at zero temperature can be 
analysed with simple energetic arguments. For small
  values of the field $H$, all spins are pointing in the same direction, say 
up, because boundary negative fields are not 
  strong enough to compete with transverse Ising couplings and reverse the 
corresponding spins (see figure \ref{configurations}a). The energy in this case 
is $E_0$. When increasing the field, the spins $\sigma_{L_y/2+1,1}\dots  
\sigma_{L_y,1}$ will eventually reverse their sign, and the corresponding energy 
is $E_1=E_0-HL_y+2J(L_y/2+2)$, which is lower than $E_0$ if $H>J(1+4/L_y)$ 
(see figure \ref{configurations}b). Another possible configuration is when all spins $\sigma_{L_y/2+1,j}\dots  \sigma_{L_y,j}$, for $j=1\dots L_x$, reverse 
their sign (see figure \ref{configurations}c). In this case, the total 
magnetisation is zero, and the corresponding energy is $E_2=E_0-HL_y+4JL_x$, 
which is  lower than $E_0$ for $H>4\zeta J$. Comparing the energies  $E_1$ and 
$E_2$, we 
conclude that the interface stays on the boundary if $\zeta>1/4+1/L_y=\raps$ 
($E_1<E_2$, and 
$H>J(1+4/L_y)$), and propagates inside the bulk when $\zeta$ is smaller than 
the critical ratio value $\raps$ and $H$ larger than $\Hs\equiv 4\zeta J$. In the latter case, where the total bulk magnetisation spontaneously goes from unity to 
zero, the transition is first order. \\
  In the thermodynamic limit, $\raps$ tends to $1/4$. This particular value is 
actually deeply related to the boundary condition in the $y$-direction. For free 
boundary conditions we would have found $\raps=1/2$ instead. The physical 
interpretation of this threshold is simplified by the study of 
boundary spin-spin correlation function for various $\zeta$. A direct extension 
of results in~\cite{jpa05} leads to the figure \ref{crossover}. For $L_x \ll 
L_y$ we observe an exponential decay of the correlation functions, typical of a 
1D behaviour. If the aspect ratio increases, we observe a obvious crossover 
towards a 2D behaviour at large $\zeta$. The crossover is obtained in this case 
for $\zeta \simeq 1/4$. \\
  In particular, we would like to know in the following how the transition line 
$\Hs(T)$ for $\zeta < 1/4$  behaves as the temperature is increased to near the second order phase 
transition at $T_\mathrm{c}/J=1/\tanh^{-1}(\sqrt{2}-1)
  \simeq 2.27$.
\section{Analysis of the thermodynamic limit}\label{analysis}
The Hamiltonian (\ref{Ham}) leads to the decomposition of the total free energy 
${\cal F}$ as follows:
\be \nn
{\cal F}(T,H)={\cal F}_0(T,H)+{\cal F}_{\mathrm{hom}}(T,H)+ \sint(T,H),
\ee
where ${\cal F}_0$ is the free energy in zero field, ${\cal F}_{\mathrm{hom}}$ the 
additional free energy corresponding to an \textit{homogeneous} boundary 
magnetic field $H$. The contribution $\sint$ is the corrective term due to the 
change of sign of boundary magnetic field. The exact expressions for those 
terms in the discrete case can be found in \cite{jpa05}. As $\sint$ is the only term 
corresponding to inhomogeneous surface conditions, it contains all the physical 
informations about the transition described in the previous section. The 
section (\ref{31}) starts with the calculation of $\sint$ in the thermodynamic limit and the evaluation of
finite size corrections as well. This allows us to characterize the 
details of the transition: In section (\ref{32}), we obtain the expression of 
the transition line and the corresponding phase diagram, and in section 
(\ref{33}) we analyse the behaviour of this line at low temperature and close to 
the bulk critical point. Finally, in section (\ref{34}), we summarize the different results
and the physical interpretation.\\
\subsection{Expression of the interfacial free energy}\label{31}
We start with the discrete expression of the boundary free energy $\sint$ taken from
reference~\cite{jpa05} (see equation (84) in that paper). It was obtained by computing
two point correlation functions with a Grassmannian quadratic action expressed in Fourier modes.
The corresponding result is the following: 
\be\label{sigmaint}
-\beta \sint=\log
\left [
1-\frac{2}{L_y}\sum_{q=0}^{L_y/2-1}(-1)^q\cot (\theta_{q+\fd}/2) 
F(\cos\theta_{q+\fd})
\right ],
\ee
where $\theta_q=(2\pi /L_y)(q+1/2)$, and $F$ is a function of $t=\tanh \beta J$ 
and $u=\tanh \beta H$ defined
as 
\be
\fl
F(x)&=&\label{functionF}
\frac{4tu^2G(x)}{
\frac{1}{4}[1-(1+t^2)(t^2+2tx-1)G(x)]^2
+2tu^2(1+x)
G(x)+4t^4(1-x^2) G(x)^2
}.
\ee
The function $G$ is defined by
\be\label{functionG}
G(x)&=&\frac{1}{L_x}\sum_{p=0}^{Lx-1}\frac{1}{
(1+t^2)^2-2t(1-t^2)(\cos\theta_p+x)},
\ee
where $\theta_p=2 \pi p/L_x$. We propose in this section to simplify the expression~(\ref{sigmaint}) 
by studying the thermodynamic limit, in order to obtain the dominant terms contributing to the free energy.
The sum inside the logarithm function 
(\ref{sigmaint}) has the particularity to 
behave like a Dirac distribution in the limit $L_y\rightarrow \infty$. Indeed, 
we define first the 
following sum 
\be\label{sumS}
S[F]=\frac{2}{L_y}\sum_{q=0}^{L_y/2-1}
(-1)^q\cot (\tq/2)F(\cos(\tq)),
\ee
for any function $F$. In \ref{DiracSum}, we demonstrate that in the 
thermodynamic limit,
this sum is simply $F(1)$, plus corrections which are exponentially small in the 
system size $L_y$. These corrections  are important for finding the asymptotic 
behaviour of the 
interface free energy. We are expecting 
$\sint$ to be linear in $L_x$ or $L_y$, so the argument
in the logarithm (\ref{sigmaint}) should be exponentially small in $L_x$ or 
$L_y$. In particular, the $L_y$ dependence
is contained in the corrections of the Dirac distribution, and the $L_x$ 
dependence
is contained in the term $F(1)$ through
the finite sum $G$, equation (\ref{functionG}).
 From \ref{DiracSum} and equation (\ref{distrib}), we can expand $S[F]$ in the 
limit of large $L_y$ and we obtain for the dominant part
 \be
 S[F]\simeq F(1)-A_{L_y/2},
 \ee
 where $A_{L_y/2}$ is the $L_y/2$-th Fourier coefficient of the function $F$, 
and it will be shown that
 it goes to zero exponentially in $L_y$.  The term $F(1)$ is equal to
 \be\nn
 F(1)=\frac{4tu^2G(1)}{[1-(1+t^2)(t^2+2t-1)G(1)]^2/4+4tu^2G(1)},
 \ee
 and $G(1)$ given by (\ref{functionG}) can be expanded as a Fourier series the 
following 
way
\be
G(1)=\frac{1}{L_x}\sum_{p=0}^{Lx}\sum_{k\ge 0}B_k\cos(k\theta_p),
\ee
where the Fourier coefficients $B_k$ are defined by
\be
B_0&=&\frac{1}{2\pi}\int_{0}^{2\pi} \dd \theta \frac{1}{
(1+t^2)^2-2t(1-t^2)(1+\cos\theta)},
\\ \nn
B_{k>0}&=&\frac{1}{\pi}\int_{0}^{2\pi}\dd \theta \frac{\cos(k\theta)}{
(1+t^2)^2-2t(1-t^2)(1+\cos\theta)}.
\ee
Using
\be\label{formula}
\frac{1}{2\pi}\int_0^{2\pi}\frac{\dd \theta}{a-b\cos\theta}=\frac{1}{
\sqrt{a^2-b^2}},
\ee
we arrive at the expression  $B_0=1/[(1+t^2)|t^2+2t-1|]$. It is then easy to 
show that
\be
G(1)=\sum_{k\ge 0}B_k\frac{1}{L_x}\sum_{p=0}^{L_x-1}\cos(k\theta_p)
=B_0+B_{L_x}+B_{2L_x}+\dots
\ee
We expand then $F(1)$ for $L_x$ large, when $B_ {L_x}$ is small, and it is 
sufficient to keep the first two terms:
\be\nn
 F(1)&\simeq&\frac{4tu^2(B_0+B_{L_x})}{
\frac{1}{4}[1-(1+t^2)(t^2+2t-1)(B_0+B_{L_x})]^2+4tu^2(B_0+B_{L_x})},
\\
&\simeq &1-\frac{(1+t^2)^2(t^2+2t-1)^2}{16tu^2}\frac{B_{L_x}^2}{B_0}.
\ee
We have check that the coefficient $B_{2L_x}$, which would be of the order of 
$B_{L_x}^2$, does not
appear at this order. $B_{L_x}$ can be computed analytically in the complex 
plane.
If we define $z=\exp(i\theta)$, we have $B_{L_x}={\rm Re}\oint 
\dd z/(2i\pi)4z^{L_x}/Q(z)$,
where $Q(z)$ is a polynomial function 
\be
Q(z)&=&-2t(1-t^2)z^2+2z[(1+t^2)^2-2t(1-t^2)]-2t(1-t^2),
\\ \nn
&=&-2t(1-t^2)(z-z_-)(z-z_+).
\ee
The zeros $z_{\pm}$ of this function are distributed on the 
positive real axis with $z_-=(1-t)/[t(1+t)],\;\;z_+=1/z_-$.
$z_-$ is less than 1 (or $z_-<z_+$) for $t>\tc=\sqrt{2}-1$, in the 
low temperature regime (therefore the quantity $t^2+2t-1$ is always positive).
The value of $B_{L_x}$ in this region is then equal to 
\be
B_{L_x}=\frac{2}{(1+t^2)(t^2+2t-1)}\left (\frac{1-t}{t(1+t)}\right )^{L_x}.
\ee
We obtain therefore, for the dominant part of the distribution $S$:
\be
S[F]\simeq 1-\frac{(1+t^2)(t^2+2t-1)}{4tu^2}
\left (\frac{1-t}{t(1+t)}\right )^{2L_x}-A_{L_y/2}.
\ee
These corrective terms are all negative, since $1-S[F]$ should be positive so
that the logarithm in (\ref{sigmaint}) is always defined, and the interfacial free energy can now be written as
\be\nn 
-\beta \sigma_{\mathrm{int}}&=&\log\left (
1-S[F] \right ),
\\
&\simeq&
\log \left (\frac{(1+t^2)(t^2+2t-1)}{4tu^2}
\left (\frac{1-t}{t(1+t)}\right )^{2L_x}+A_{L_y/2}
\right ).
\ee
The Fourier coefficient $A_{L_y/2}$ is evaluated for large $L_y$ with the 
function (\ref{functionF}). We expect this coefficient to be exponentially
small with $L_y$, with some corrections to this coefficient which are also small 
in $L_x$. Therefore the dominant term can be obtained by taking the limit 
$L_x=\infty$
in (\ref{functionG}). Using the formula (\ref{formula}) we obtain
in this limit 
\be \fl
G(\cos\theta)&=&\frac{1}{\sqrt{R(\cos\theta)}},
\\ \nn \fl
R(\cos\theta)&=&[(1+t^2)^2+2t(1-t^2)(1-\cos\theta)][(1+t^2)^2-2t(1-t^2)
(1+\cos\theta)].
\ee
The function $F$ can then be rewritten, after some algebra, as 
\be \fl
F(\cos\theta)=\frac{8tu^2}{4tu^2(1+\cos\theta)+(1+t^2)(1-2t\cos\theta-t^2)
+\sqrt{R(\cos\theta)}}.
\ee
Now $A_{L_y/2}$ can be expressed by mean of a complex 
integration along the unit circle
\be\nn
A_{L_y/2}&=&{\rm{Re}}\frac{1}{\pi}\int_0^{2\pi}\dd \theta F(\cos\theta)\exp\left(i\frac{L_y}{2}
\theta \right),
\\ 
&=&{\rm{Re}}\oint \frac{\dd z}{2i\pi}F\left (\frac{z^2+1}{2z}
\right )2z^{L_y/2-1}.
\ee
The value of this integral depends on the poles of the function $F$. 
Setting $X=(z^2+1)/2z$, we can write
\be
F(X)=\frac{8tu^2}{P(X)+\sqrt{R(X)}}=\frac{8tu^2\Big(P(X)-\sqrt{R(X)}\Big)}{
P(X)^2-R(X)}.
\ee
The polynomial $P(X)^2-R(X)$ is a second order polynomial in $X$, and has two 
zeros which are
\be
X_0=\frac{1}{2}\frac{2t^3-2tu^4-u^2(1-t^4)}{t(1-u^2)(t^2-u^2)},\;\;
X_1=-1.
\ee
$X_1$ is not a pole of the function $F$ since it is not a zero
of $P(X)+\sqrt{R(X)}$. Indeed, we have $P(-1)+\sqrt{R(-1)}=(1+t^2)(
1+2t-t^2+|1+2t-t^2|)>0$, because $1+2t-t^2$ is always positive in the
interval $0<t<1$.
Then only $X_0$ is pole of $F$, and in the complex plane this gives
two solutions $Z_{\pm}$ of the equation $z^2-2X_0z+1=0$.
If $Z_-$ is the solution which is inside the unit circle ($|X_0|>1$), the 
value of $A_{L_y/2}$ is given by
\be
A_{L_y/2}&=&{\rm Re}\left ( 
\frac{P(X_0)-\sqrt{R(X_0)})}{
(P(X)^2-Q(X))'(X_0)\sqrt{X_0^2-1}}Z_-^{L_y/2-1}
\right ),
\\ \nn
&\equiv& C_y(t,u)|Z_-|^{L_y/2}.
\ee
In some cases the zeros can be on the unit circle ($|X_0|<1$), but we assume 
this latter situation is 
not physical since this would mean that the thermodynamic limit is never 
obtained.\\
Finally the free energy for the interface in the large system size limit 
can be written as the logarithm of two dominant terms
\be\label{resultsigma} \fl
-\beta \sint\simeq
\log \left [\frac{(1+t^2)(t^2+2t-1)}{4tu^2}
\left (\frac{1-t}{t(1+t)}\right )^{2L_x}+C_y(t,u)|Z_-|^{L_y/2}
\right ].
\ee
\subsection{Line transition and phase diagram}\label{32}
\begin{figure}[!t]
\begin{center}
\includegraphics[scale=0.4]{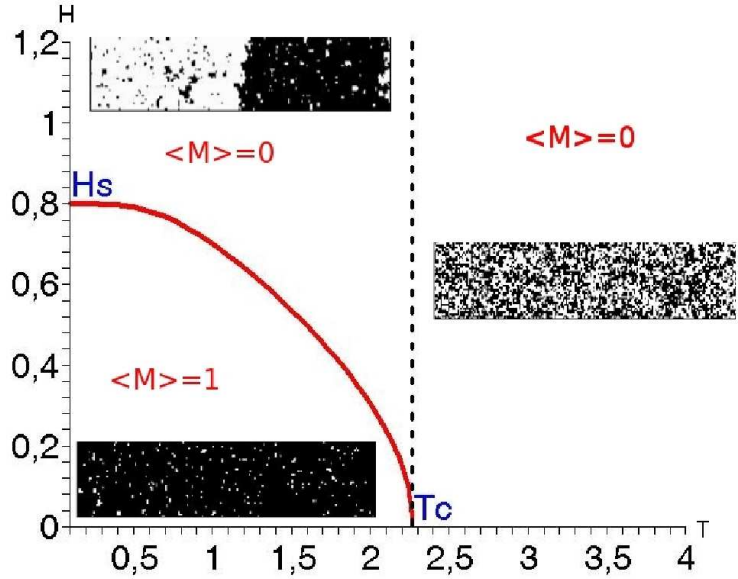}
\caption{\label{phasediag}Phase diagram for the system at $\zeta=0.2$. The plain 
line represents the first order transition given by equation \eref{criticalline}. 
The dashed line is the bulk 2$^\mathrm{nd}$ order phase 
transition. The snapshots are extracted from Monte Carlo simulations in the 
different regimes, for $L_x=40$ and $L_y=200$.}
\end{center}
\end{figure}
Two different regimes can be identified from equation (\ref{resultsigma}), depending 
on whether the first 
term in the logarithm is greater or smaller than the second one. In the first case, the free energy is proportional to $L_x$ and 
a coefficient which does not depend on magnetic field, only on $t$. 
At zero temperature, it is easy to show that $\sint\simeq 4J\zeta L_y$, which 
corresponds to the energy $E_2$ computed above in section~(\ref{model}). This means that this term 
represents a situation where it is energetically favourable for the interface to 
spread inside the bulk. On the other hand, the second term in the logarithm of 
equation ($\ref{resultsigma}$) represents a configuration where the interface is 
localised on the boundary. Between these two regimes, we can define a line of transition which is \textit{a priori} first order: The first term does not depend on the magnetic field, therefore the free energy has in 
general a cusp as a function of $u$) by making the two exponential terms in 
(\ref{resultsigma}) equal in magnitude.\\
We then obtain the transition line equation in the $(t,u)$-plane which is 
simply a quartic polynomial in $u$:
\be\label{criticalline} \fl
2t\Big(1+v(4\zeta)\Big)u^4+(1+t^2)\Big(1-2tv(4\zeta)-t^2\Big)u^2+2\Big(v(4\zeta)-1\Big)t^3=0,
\\ \nn
v(4\zeta)=\cosh \left [ 4\zeta\ln \left (\frac{1-t}{t(1+t)}\right ) \right ].
\ee
On figure \ref{phasediag} is shown the phase diagram for the system at $\zeta=0.2$,
with snapshots of typical configurations at low and large 
fields/temperatures, obtained by Monte Carlo numerical simulations.

\subsection{Behaviour around $T=0$ and $T=T_c$}\label{33}
Near zero temperature, we can expand the previous relation with
$t=(1-\epsilon)/(1+\epsilon)$, $\epsilon=\exp(-2/T)\ll 1$. We obtain
at lowest order in $\epsilon$ and for $4\zeta<1$:
\be\label{expansionU}
(1+2\epsilon^{4\zeta})u^4-2u^2+1-2\epsilon^{4\zeta}=0,
\ee
which gives, $u^2=1$ or $u^2=(1-2\epsilon^{4\zeta})/(1+2\epsilon^{4\zeta})$.
The non trivial solution gives the point $\Hs=4\zeta J$ as expected at zero 
temperature from preliminary study. Moreover, if the transition line ends at the point $u=0$, this 
is equivalent to $X_0=v=1$, or $1-t=t(1+t)$, which gives 
$t=\tc$.  The line ends therefore at the second order transition point. In this 
case $Z_{+}=Z_{-}=1$, which is the transition value between an exponential 
behaviour and oscillating one in the logarithm arguments (\ref{sigmaint}). This 
basically suggests that the interface free energy is no more an extensive 
function of the system size, and does not contribute to the thermodynamic 
behaviour of the system. In the case when $4\zeta>1$, the discriminant $\Delta$ 
of the equation (\ref{criticalline}) can be written as
\be\nn \fl
\Delta&= &4t^2(1-t^2)^2\left (
v-\frac{(1+t^2)^2+2t(1-t^2)}{2t(1-t^2)}\right )
\left (
v-\frac{(1+t^2)^2-2t(1-t^2)}{2t(1-t^2)}\right ),
\\ \fl
&=&4t^2(1-t^2)^2\left (
v-\frac{(1+t^2)^2+2t(1-t^2)}{2t(1-t^2)}\right )
\left (
v-v(1)\right ).
\ee
Expanding $v(4\zeta)$ near the threshold value $4\raps=1$, we obtain 
\be\nn \fl
v(4\zeta)\simeq v(1)+(4\zeta-1)\ln \left [\frac{1-t}{t(1+t)}\right ]
\sinh \left (\ln \left [\frac{1-t}{t(1+t)}\right ] \right ),
\ee
and the discriminant can be expanded as
\be \label{Discriminant}\fl
\Delta \simeq 8t^2(1-t^2)^2\ln \left [\frac{1-t}{t(1+t)}\right ]
\sinh \left (\ln \left [\frac{1-t}{t(1+t)}\right ] \right ) (1-4\zeta).
\ee
The discriminant $\Delta$ is negative when $\zeta>1/4$ and equation (\ref{criticalline}) has 
no real solution, therefore the wetting transition no longer exists in this 
regime.\\
Near zero temperature and for $\zeta<1/4$, we can expand more precisely 
equation (\ref{criticalline}), for small parameter $\epsilon$,
this will give locally the behaviour of $\Hs(T)$ as function of temperature. 
We obtain for the quantity $v$:
\be \nn
v= \fd \epsilon^{-4\zeta}(1-8\zeta\epsilon+\epsilon^{8\zeta}+\dots).
\ee
This has to be equal to the expansion of $X_0$, which is given, at lowest 
order, by
\be \nn
X_0= \fd \epsilon^{-H}(1+\epsilon^{2H}+\dots).
\ee
By comparing the two previous expressions, we obtain $\Hs(T)\simeq 4\zeta-4\zeta 
T\exp(-2/T)$. Figure \ref{phasediag} shows that, as expected, the curve 
$\Hs(T)$
is flat near zero temperature due to exponentially small thermal 
excitations.
 Near the critical temperature $t_\mathrm{c}$, a simple analysis 
gives $\Hs(T)\propto \sqrt{t-t_\mathrm{c}}$. %

\subsection{Summary of the results}\label{34}
To summarize the previous calculations, we have found that the free energy~(\ref{sigmaint}) can be 
expressed as the logarithm of two contributions which are exponentially small in, respectively,
the system sizes $L_x$ and $L_y$, see equation~(\ref{resultsigma}). This result is 
obtained after performing an asymptotic analysis of the Fourier sum $S$, equation (\ref{sumS}), that
appears in the logarithm of equation~(\ref{sigmaint}). This sum behaves like a Dirac distribution in the 
thermodynamic limit (see~\ref{DiracSum}) 
and its value in this limit makes the logarithm in~(\ref{sigmaint}) singular. To remove the singularity we 
analysed the finite size dependent corrective terms since we expect $\sint$ to diverge linearly with the system size.  
The two main contributions inside the logarithm essentially come from two relevant Fourier amplitudes that tend to zero 
exponentially with $L_x$ and $L_y$.
The other contributions are much smaller and do not contribute to the free energy. One term corresponds to the interface
localized on the boundary and the other to the interface extended across the bulk. The relative amplitude between these two terms
is controlled by the aspect ratio $\zeta$, and the transition line between the two regimes is expressed by a simple quartic equation 
(\ref{criticalline}). A study of its discriminant~(\ref{Discriminant}) leads to the existence of a first order transition line when $\zeta<1/4$. For $\zeta>1/4$ no real solution exists, and the interface is always localized on the boundary since only one of the two contributions inside the logarithm is dominant at all temperatures and magnetic fields.
  
\section{Conclusion}
In this article we obtained an exact description of a first order phase 
transition induced by a simple inhomogeneous boundary magnetic field. The use of 
Grassmann techniques allows for an exact calculation of the interfacial free 
energy in the discrete case, which is
suitable to study then the thermodynamic limit by asymptotic methods 
presented in this paper. This approach allows us to control the way the 
thermodynamic limit is taken, and has the advantage that no cut-off parameter is 
required for the continuum limit. In particular, we have demonstrate that it is 
straightforward to take the thermodynamic limit exactly for a given geometry. 
This leads to a surprisingly simple equation of the transition line (\ref{criticalline}) in the $(H,T)$ 
or $(u,t)$ planes, and the corresponding critical behaviour close to the bulk critical point $(0,T_c)$. In the context of wetting 
transitions this is a non trivial extension of previous results. This line 
disappears for $\zeta>1/4$ as the solutions move to 
the complex plane. A numerical study of bulk correlation functions at the 
precise value $\zeta=1/4$ might show the dynamical 
instability of the interface on the vanishing transition line. The infinite strip 
models for inhomogeneous surface field~\cite{abraham80,abraham88,abraham81} 
might not capture this feature since the interface is not sensitive to the
system aspect ratio.\\
Grassmann techniques, in complement to conformal theory~\cite{chatterjee,mussardoBMF} and transfer matrix 
methods, can be considered as an interesting optional way to solve 
boundary problems or wetting transitions. 
Extensions of the method presented here and in~\cite{jpa05} might be useful to study other 
kind of wetting transitions. In
particular it might be applied to models of defects other than a surface field 
such 
as a line of weaker or stronger couplings~\cite{fisher67,bariev79}, which has 
been studied in the framework of transfer matrix methods in detail, and where 
striking similarities with other physical domains like electrostatics~\cite{ko85} 
have been suggested. 
\appendix

\section{}\label{DiracSum}
We would like to compute the following sum in the large $L$ (even) limit ($L_y$ 
is replaced by $L$ here for generality):
\be
S[F]=\frac{2}{L}\sum_{q=0}^{L/2-1}
(-1)^q\cot (\tq/2)F\Big(\cos\tq\Big),
\ee
with $\tq=(2\pi)/L(q+\fd)$, and $F$ any function of the variable
$\cos \tq$. We know that for any constant $F$, $S[F]=F$ (see \ref{Sum1}).
For commodity, we can extend the sum from $q=0..L/2-1$ to $q=0..L-1$ by writing 
\be\nn \fl
S[F]=\fd\frac{2}{L}\sum_{q=0}^{L/2-1}
(-1)^q\cot \left[\frac{\pi}{L}\left(q+\fd 
\right)\right]F\left[\cos\left(\frac{\pi}{L}\left(q+\fd\right)\right)\right]
\\
\nn \fl
+\fd\frac{2}{L}\sum_{q=L/2}^{L-1}
(-1)^{L-q-1}\cot 
\left[\frac{\pi}{L}\left(L-q-1+\fd\right)\right]F\left[\cos\left(\frac{2\pi}{L}\left(L-q-1+\fd\right)\right)\right].
\ee
It is then straightforward to see that, after rearranging the different terms,
\be
S[F]=\frac{1}{L}\sum_{q=0}^{L-1}
(-1)^q\cot \left[\frac{\pi}{L}\left(q+\fd 
\right)\right]F\left[\cos\left(\frac{\pi}{L}\left(q+\fd\right)\right)\right].
\ee
Next we express the function $F$ as a Fourier series:
\be
F(\cos\theta)=\sum_{p=0}^{\infty}A_p\cos(p\theta).
\ee
Computing $S[F]$ is equivalent to obtain explicitly the value of 
every term $S[\cos(p\theta)]$. For $p=1$, we have
\be\nn \fl
\cot\left(\fd \tq\right)\cos\tq&=&\frac{1+\cos\tq}{\sin\tq}\cos\tq
=\frac{\cos\tq+\cos^2\tq}{\sin\tq},
\\ \nn \fl
&=&\frac{\cos\tq+1}{\sin\tq}-\sin\tq,
\ee
which implies that 
\be
S[\cos(.)]=S[1]-\frac{1}{L}\sum_{q=0}^{L-1}(-1)^q\sin\tq.
\ee
For the second term, $p=2$, we can show that
\be\nn \fl
\cot\left(\fd \tq\right)\cos(2\tq)&=&\frac{1+\cos\tq}{\sin\tq}\cos(2\tq),
\\ \nn \fl
&=&\frac{\cos\tq+1}{\sin\tq}\cos\tq-\sin(2\tq)-\sin\tq,
\ee
and for the third term
\be\nn \fl
\cot\left(\fd \tq\right)\cos(3\tq)&=&\frac{1+\cos\tq}{\sin\tq}\cos(3\tq),
\\ \nn \fl
&=&\frac{\cos\tq+1}{\sin\tq}\cos(2\tq)-\sin(3\tq)-\sin(2\tq).
\ee
By recursion, it is easy to show that
\be
S[\cos(p.)]&=&S[\cos([p-1].)]-T_p-T_{p-1},\\
\nn
T_p&=&\frac{1}{L}\sum_{q=0}^{L-1}(-1)^q\sin(p\tq).
\ee
Finally, after rearranging the previous relations, we obtain the following 
expression 
\be
S[\cos(p.)]&=&S[1]-T_p-2(T_{p-1}+T_{p-2}+...+T_1),\;\;S[1]=1,
\\ \nn 
S[\cos(.)]&=&S[1]-T_1.
\ee
The quantities $T_p$ can be easily computed, and we obtain
$T_p=0$ except for $p=L(\fd+k)$, $k\ge 0$, where $T_p=(-1)^k$.
After some combinatorial, the distribution $S[F]$ is equal to
\be\label{distrib}
S[F]=F(1)-\sum_{k=0}^{\infty}A_{L/2+kL}-2\sum_{k=0}^{\infty}\sum_{k'=1}^{L-1}
A_{L/2+2kL+k'}.
\ee
In the limit where $L$ is infinite, $S[F]$ is the Dirac distribution
$S[F]=F(1)$ since all the Fourier coefficients tend to zero. 
\section{}\label{Sum1}
In this section, we want to prove the following equality
\be
S[1]=\frac{1}{L}\sum_{q=0}^{L-1}
(-1)^q\cot \left[\frac{\pi}{L}(q+\fd)\right]=1,
\ee
for any value of $L$.  Let assume that $L$ is even, the proof for $L$ odd is 
equivalent.
We can notice that
\be
S[1]=\frac{1}{L}\sum_{q=0}^{L-1}
(-1)^q\frac{\partial}{\partial z}\log \left |\sin \left 
[\frac{\pi}{L}(q+\fd)+z\right ]\right |_{z=0}.
\ee
Separating in the sum the odd and even integers $q$, we obtain easily 
\be
S[1]=\frac{1}{L}\frac{\partial}{\partial z}
\log 
\left |
\prod_{q=0}^{L/2-1}\frac{\sin [\pi(2q+1/2)/L+z]}{\sin [\pi(2q+3/2)/L+z]}
\right |_{z=0}.
\ee
The 2 products  are evaluated after expressing the sine function as exponential 
terms, and 
using by identification the equality  $X^{L/2}-1=\prod_{q=0}^{L/2-1}\left 
[X-\exp(4i\pi q/L)\right ]$. We then obtain
the simple result
\be
S[1]&=&\frac{1}{L}\frac{\partial}{\partial z}
\log 
\left |
\frac{\sin (zL/2+\pi/4)}{\sin (zL/2+3\pi/4)}
\right |_{z=0},
\\
&=&\fd \left[\cot (\pi/4)-\cot (3\pi/4)\right]=1.
\ee
By extension, $S[C]=C$ for any constant $C$.
\bibliographystyle{unsrt}

\end{document}